\documentstyle[epsfig,amssymb]{mn}
\newcommand{\etal}{{\it et al. }}
\newcommand{\uv}{{\sl uv }}
\begin{document}

\title[Jet reorientation in AGN]{Jet reorientation in AGN: two winged
radio galaxies} 
\author[J. Dennett-Thorpe et al.]
{J. Dennett-Thorpe,$^{1}$ \thanks{jdt@astro.rug.nl} P.~A~.G.
Scheuer$^2$, R.~A.~Laing,$^{3,4}$ A.~H.~Bridle,$^5$ \newauthor G.~G. Pooley,$^2$
W.~Reich$^6$ \\
$^1$ Kapteyn Institute, Postbus 800, 9700 AV Groningen, NL\\      
$^2$ Mullard Radio Astronomy Observatory, Cavendish
     Laboratory, Madingley Road, Cambridge CB3 0HE\\ 
$^3$ Space Science and Technology Department, CLRC, Rutherford
     Appleton Laboratory, Chilton, Didcot, Oxfordshire OX11 0QX \\
$^4$ University of Oxford, Department of Astrophysics, 
     Nuclear and Astrophysics Laboratory, Keble Road, Oxford OX1 3RH \\
$^5$ NRAO, 520 Edgemont Rd, Charlottesville, VA 22903-2475, USA \\ 
$^6$ MPIfR, Auf dem H{\"u}gel 69, D-53121 Bonn, Germany\\ }

\date{Received }
\maketitle

\begin{abstract}

Winged, or X-shaped, radio sources form a small class of
morphologically peculiar extragalactic sources. We present
multi-frequency radio observations of two such sources. We derive
maximum ages since any re-injection of fresh particles of 34 and
17\,Myr for the wings of 3C\,223.1 and 3C\,403, respectively, based on
the lack of synchrotron and inverse Compton losses.  On morphological
grounds we favour an explanation in terms of a fast realignment of the
jet axis which occurred within a few Myr. There is no evidence for
merger activity, and the host galaxies are found to reside in no more
than poor cluster environments. A number of puzzling questions remain
about those sources: in particular, although the black hole could
realign on sufficiently short timescales, the origin of the
realignment is unknown.
\end{abstract}


\section{Introduction}

`Winged' or `X-shaped' extragalactic radio sources are notable for
their peculiar radio morphologies, which may indicate realignments of
the axes of their radio jets. These sources have large scale extrusions
of radio plasma apparently beginning close to the core, which are
about the same length as, or longer than, the presently active lobes.
Determining if jet reorientation is required, and the timescales on
which any realignment takes place provides information which may be
used in understanding the process of jet formation. In particular we
may address whether the jet axis is related to the spin-axis of the
black hole, or that of the accretion disc.

Possible mechanisms for the formation of these objects include
expansion of old, differently aligned lobes; backflow into the cavity
of such lobes; and relatively slow conical precession of the jet axis,
resulting in the observed x-shaped morphology by projection. These
models will be dealt with in more detail in section 5. 

In this paper we present multi-frequency, high-resolution,
high-sensitivity radio observations of two such `winged' sources --
3C\,223.1 and 3C\,403 -- in an attempt to constrain their manner of
formation, and the timescales involved. These sources are two of the 7
least luminous `FRII' sources in the 3CR catalogue
\cite{ben62,spi85}. We note that all known sources with this
morphology show radio luminosities close to the FRI/FRII break
(10$^{25}$W\,Hz$^{-1}$ at 178\,MHz; Fanaroff \& Riley (1974)).

As both sources discussed here are narrow--line radio galaxies,
the assumptions of simple Unified Schemes (Scheuer 1987; Barthel
1989), as well as the length/width ratio of the lobes and wings,
require that the large linear size of the wings ($\sim$ twice that of
the active lobes) cannot be explained as the projection effect of a
small excrescence.

We discuss the observations in section~\ref{sec:obs}. We describe the
morphology of the sources, including their polarization structures
(\ref{sec:morph}), consider the distribution of the radio spectra across
the sources, and derive maximum particle ages (\ref{sec:spect}) and
examine their rotation measures and (de)polarization properties
(\ref{sect:RM}). We briefly comment on other possibly related sources in
section~\ref{sec:comp}. We compare our radio data with the expectations of
different formation scenarios in section~\ref{sec:form}. Finally, we
comment on the possibilities of achieving the required timescales for
reorientation of the jet axis, given observations of the host galaxies
in the literature (\ref{sec:merge}) and recent theoretical advances
(\ref{sec:realign}).

Throughout the paper we assume a q$_0$=0.5, $\Lambda$=0 cosmology with
H$_0$ = 65\,km\,s$^{-1}$Mpc$^{-1}$. 

\begin{table*}
\caption{Basic source parameters}
\label{tab:basic}
\begin{tabular}{lrrlrrrrrr}
\hline
\hline
Source & RA (J2000) & dec (J2000) & z & S$_{178MHz}$& log(P$_{178}$)&lobe&lobe& largest angular&wing projected\\
       &            &             &   & (Jy)     & (W\,Hz$^{-1}$) &
       (arcsec) & (kpc)& size (arcsec) & size (kpc)\\
\hline
3C\,223.1& 09 41 23.92 & +39 44 41.7 & 0.1075 & 8.1  & 25.29 &85&170&170 & 340\\
3C\,403  & 19 52 15.29 & +02 30 28.0 & 0.059  & 28.3 & 25.30 &110&130&230 & 275\\
\hline
\end{tabular}

S$_{178}$ is on the scale of Roger, Bridle \& Costain (1973) using
conversion factors from Laing \& Peacock (1980). The value for
3C\,223.1 has been interpolated from flux densities at neighbouring
frequencies quoted in Kellermann, Pauliny-Toth \& Williams (1969),
adjusted to the present scale. The value for 3C\,403 is the 3CR value,
corrected to the present scale, as there is no 4CT flux density. (The
value quoted by Kellermann, Pauliny-Toth \& Williams (1969) is in fact
the 4C value, and is clearly affected by partial resolution.) Redshift
for 3C\,223.1 is from Sandage (1966) and 3C\,403 from Sandage
(1972). P$_{178}$ is calculated using spectral index $\alpha$=0.67
\nocite{rog73,lai80b,san66,san72,kel69}
\end{table*}
\section{Observations}
\label{sec:obs}
\subsection{Previous observations at radio wavelengths}

High resolution maps at 8~GHz have been presented by Black \etal
\shortcite{bla92a} [B92]. At 2.5\,arcsec resolution jets are detected in the
north and east lobes of 3C\,223.1 and 3C\,403 respectively.  Higher
resolution (0.25\,arcsec) maps reveal complex hotspots (B92). The
counterjet-side hotspot of 3C\,403 has a more extended, simple structure
in comparison to the series of high brightness compact features on the
jet side. 3C\,223.1 contains large resolved hotspots in both active lobes
with similar peak surface brightnesses.  Both sources vary in width
along the length of the active lobes. This is particularly prominent in
3C\,403 where the bright knot coincides with a sharp pinching in of the
lobe.

\subsection{Radio data}
\label{sec:radobs}

3C\,223.1 was observed with the NRAO VLA at three frequencies, 3C\,403
at two.  New observations were combined with archival data, in order
to create datasets with as much \uv coverage and as much \uv overlap
between frequencies as possible.  Both sources were observed with the
Effelsberg 100m Telescope at 32\,GHz. 3C\,223.1 was also observed with
the Ryle Telescope at 15\,GHz. (The low declination of 3C\,403 make it
unsuitable for observations with this east-west array.)  All flux
densities were adjusted to the Baars et al (1977) scale, using
standard calibrators. Polarization position angles were obtained using
assumed polarization PA 33$^\circ$ of 3C\,286. Table~\ref{tab:obsrad}
lists the observing parameters.

\begin{table}
\caption{Radio observations}
\begin{tabular}{lclrrr}
\hline
\hline
date & tel. & obs. & freq & bandwidth &duration\\
      &          &          & (GHz) &(MHz) & (mins)\\
\hline
\multicolumn{6}{c}{3C223.1}\\
\hline
01 \& 15/08/82 & B & SS & 1.4 & 50 & 118\\
26/11/94 & C & JDT & 1.4 & 50 & 40\\

01 \& 15/08/82 & B & SS & 4.9 & 50 & 62\\
19/04/83 & C & SS & 4.9 & 50 & 104\\
11/04/95 & D & JDT & 4.9 & 50 & 50\\

25/05/90 & A & AB & 8.4 & 25 & 114\\
31/03/89 & B & AB & 8.4 & 25 & 88\\
02/07/89 & C & AB & 8.4 & 50 & 11\\
26/11/94 & C & JDT & 8.4 & 50 & 136\\
24/11/89 & D & AB & 8.4 & 50 & 10\\
11/04/95 & D & JDT & 8.4 & 50 & 50\\
04/95 -- 02/96& Ryle& GP& 15.2 &68.3 &96hr \\
16--17/04/97& Effel &WR& 32 & 2000 &27x15\\
\hline
\multicolumn{6}{c}{3C\,403}\\
\hline
10/06/94 & B & JDT & 1.4 & 25 & 140\\
27/11/94 & C & JDT & 1.4 & 50 & 38\\
02/07/89 & C & AB & 8.4 & 50 & 25+23\\
27/11/94 & C & JDT & 8.4 & 50 & 37+37\\
31/12/89 & D & AB & 8.4 & 50 & 9+8\\
11/04/95 & D & JDT & 8.4 & 50 & 27+24\\
17/04/97;21/05/97& Effel &WR& 32 & 200 &15x15\\
\hline
\end{tabular}
Telescope: refers to VLA array where applicable\\
Observers: AB = A. Black; JDT = J. Dennett-Thorpe; GP = G. Pooley; WR
=W. Reich; SS = S. Spangler.
\label{tab:obsrad}
\end{table}

The VLA observations were reduced entirely within {\sc aips}. The flux
densities were calibrated using the standard VLA calibrators, and
secondary calibrators observed between the target sources. The data
from a single array were both externally and self-calibrated before
being combined with data from other arrays.  When combining the data,
progressively more compact configurations were added, with further
iterations of self-calibration at each stage.

Care was taken to ensure that the observations in the synthesis arrays
at different frequencies were matched as closely as possible in their
{\sl uv} coverage. To achieve this, observations at different
frequencies had identical minimum and maximum {\sl uv} limits and
Gaussian {\sl uv} tapers applied during the imaging process in order
to generate similar synthesised beams. This procedure then
compensates for differences in density of {\sl uv} coverage near the
outside of the {\sl uv} plane. We favoured the use of the same minimum
{\sl uv} limits at all frequencies, rather than the designation of a
(poorly known) `zero-spacing' flux density.  The images produced by
imposing inner {\sl uv} limits and {\sl uv} tapers were
insignificantly different from those obtained by the cruder method of
an outer {\sl uv} cut-off only.

The extended low-luminosity wings are of great interest to us and
therefore extra care had to be taken with the imaging
procedure \cite{jdt_thesis}. The final images were made by {\sc
clean}ing with the {\sc aips} routine {\sc imagr}, and extensive
testing ensured that these results reproduced those using maximum
entropy methods. Simply comparing the total flux density in the {\sc
clean} components with that in the source (in these or in single dish
observations) is an inadequate measure of image fidelity for our
purposes. We therefore compared (as a ratio of images) the image with
the {\sc clean} components removed with an image of the {\sc clean}
components alone, paying close attention to the low surface brightness
wings. We required that $>$90\% of the flux should be in {\sc clean} components
everywhere in the source above the 6$\sigma$ contour. ($>$99\% of the total
flux density and $>$95\% of the wings' flux density is in {\sc clean}
components.)

 The large size of 3C\,403 made it necessary to use two pointing
centres at 8.4 GHz, one centred on each hot-spot. The durations in
table~\ref{tab:obsrad} refer to these separate pointings. The final
image was produced by combining in the image plane using the {\sc aips}
task {\sc vtess}.    The final effective \uv coverage will therefore not
 be the same as at 1.4\,GHz.  Tests with 3C223.1 showed, however, that
similar differences in \uv coverage produced $<$1\% changes in flux density,
and we do not believe that there is cause for concern.

The Ryle Telescope observations were performed in eight 12\,hour
periods in 1995 and 1996. The baseline coverage is from 900 to 58000
wavelengths. To obtain good \uv coverage the observations were
conducted in 4 configurations which included baselines from 1800 to
58000 wavelengths in 32 equal steps and also in one compact
configuration which included baselines 0.5, 1.5 and 2.5 times 1800
wavelengths.  The data were initially reduced using the dedicated {\sc
postmortem} package, before being transferred into {\sc aips} for
self-calibration and imaging. The Ryle Telescope observes Stokes I+Q
(figure~\ref{fig:3C223.1RT}), whilst we require total intensity
(Stokes I) only for fitting procedures in section ~\ref{sec:spect}. We
have calculated the necessary correction factors using the 8.4\,GHz
data: a good estimation as the RM for this source was only 3$\pm$4\,rad\,m$^2$ \cite{sim81a}. We took the 8.4\,GHz VLA images in total
intensity and Stokes Q (convolved to the resolution of the Ryle
image), and from this calculated an estimation of the correction
factor I/(I+Q) in each region. This correction factor was then applied
to the measured flux densities at 15\,GHz, to give the required Stokes
I flux densities. The correction factors were: 0.995 for the entire
source; 0.94, 1.14 for the S and N lobes respectively; 0.99, 0.87 for
the E and W wings respectively.

The Effelsberg data were reduced using the standard NOD 2 based
software package for `software beamswitching observations'
\cite{mor86}.  The individual maps have been combined using the {\sc
plait} algorithm by Emerson and Gr{\"a}ve \shortcite{eme88}, which
simultaneously de-stripes a set of maps observed at different scanning
directions.  Primary flux calibration was performed using 3C\,286
assuming a flux density of 2.1\,Jy. The observations were taken in
raster-scan mode, with multiple passes over the source.  The weather
throughout the observations of 3C\,223.1 was significantly worse than
that during the 3C\,403 observations.  All 15 scans were used in
3C\,403, but 5 scans were dropped from the 27 observations of
3C\,223.1 due to high noise levels. Single pixel noise spikes were
removed by hand from each of the scans. The estimation of the noise
levels in the final images is complicated both by the rotation of the
sky in the Alt-Az plane during the course of the observations (which
casues fewer scans to contribute at the edges of the final image), and
by the slightly subjective nature of the removal of noise spikes. We
have estimated the noise from off-source regions of the images well
away from their edges.  The quasi-on-off scanning technique used for
the single-dish observations can introduce small zero-level offsets
into the final image.  From the positive value of the noise
surrounding our images post-reduction, we estimate that there was a
positive zero-level of $\approx$ 5\,mJy in both cases. We have
therefore subtracted this from both images, and included this in our
error estimates.

\begin{table}
\caption{Image parameters}
\begin{tabular}{lllll}
  \hline 
  \hline 
source & freq &uv limits& beam & noise \\ 
  &(GHz)&(k$\lambda$)&(arcsec)&(mJy/beam)\\ 
\hline 
3C\,223.1& 1.4 &1.0 -- 20.0 &10.0&0.37\\ 
& 4.8 & 1.0 -- 20.0 & 10.0& 0.16\\ 
& 8.4 & 1.0 -- 20.0 &10.0& 0.13\\ 
& 32 & --& 27.0& 3.5\\ 
3C\,403 & 1.4 &0.75 - 40.0 &4.5 & 0.15 \\ 
& 8.4 & 0.75 - 40.0&4.5 & 0.11\\ 
& 32 & --& 27.0 & 5.0 \\ 
\hline
\end{tabular}
\label{tab:imagprop}
\end{table}

\section{Results}
\label{sec:results}
\subsection{Morphology and polarization structure}
\label{sec:morph}

\begin{figure*}
\begin{tabular}{cc}
3C\,223.1 &  3C\,403\\
(a)\epsfig{file=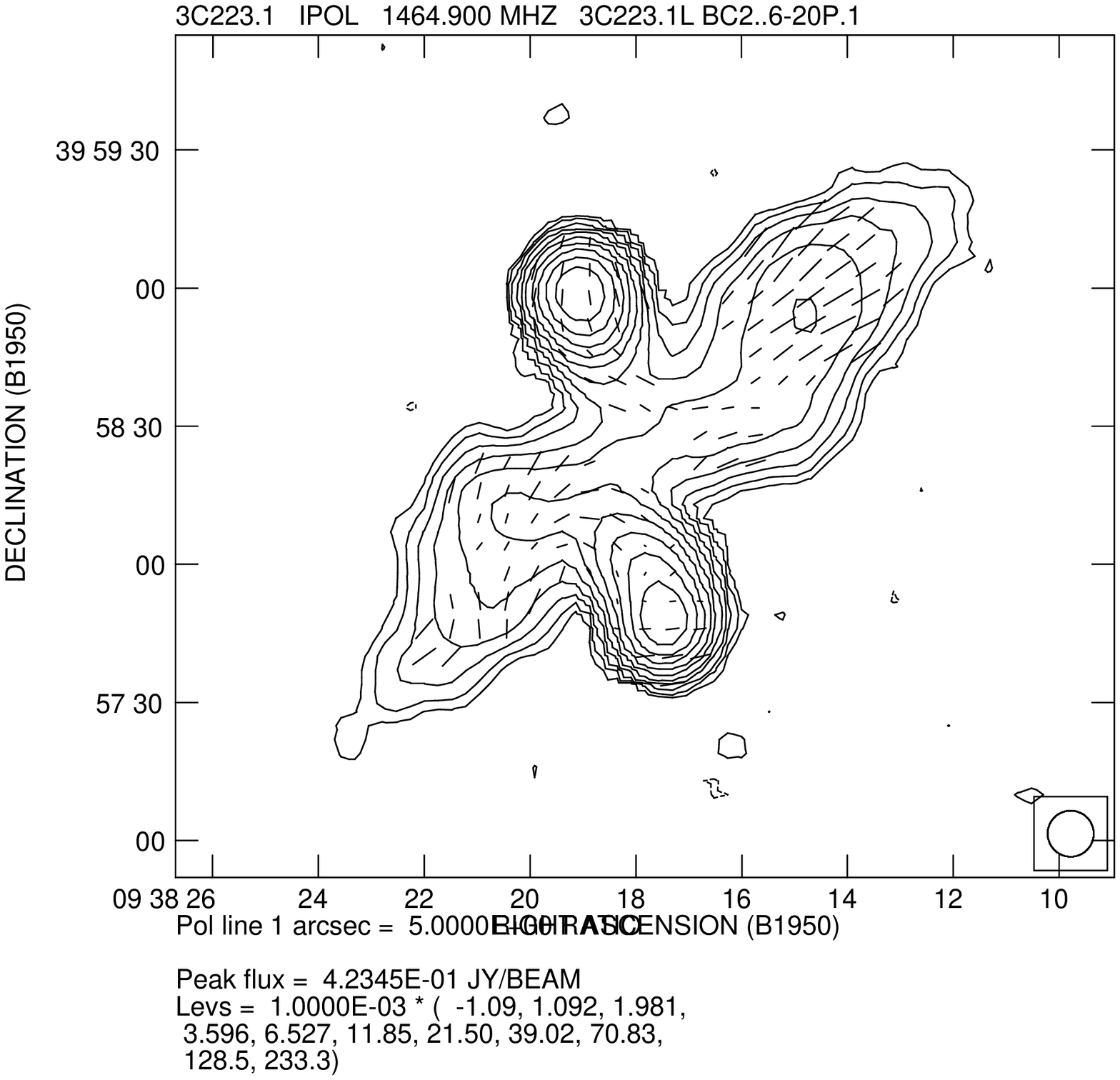,height=7.5cm,angle=-90,clip=}&
(b)\epsfig{file=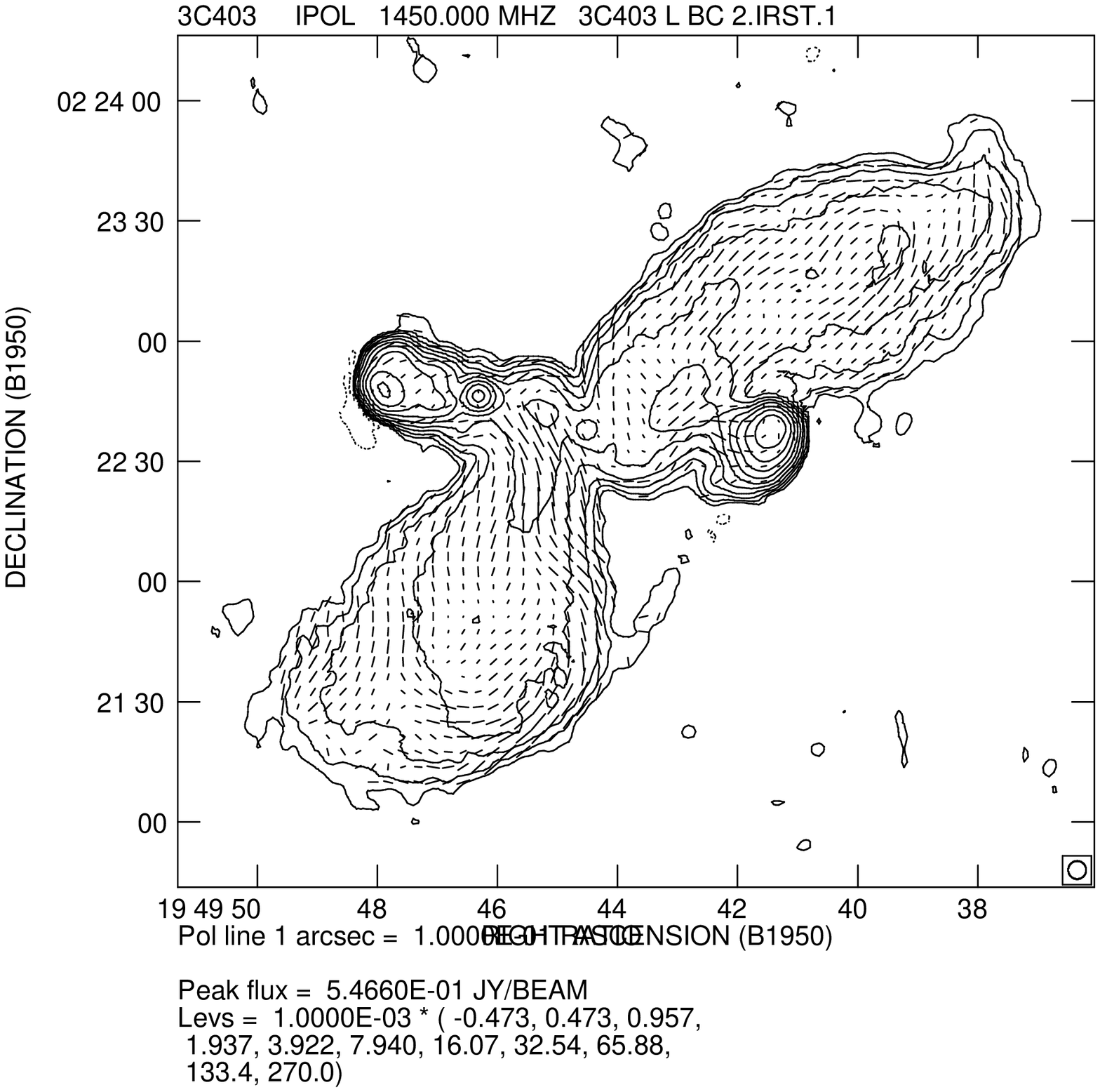,height=7.5cm,angle=-90,clip=}\\
(c)\epsfig{file=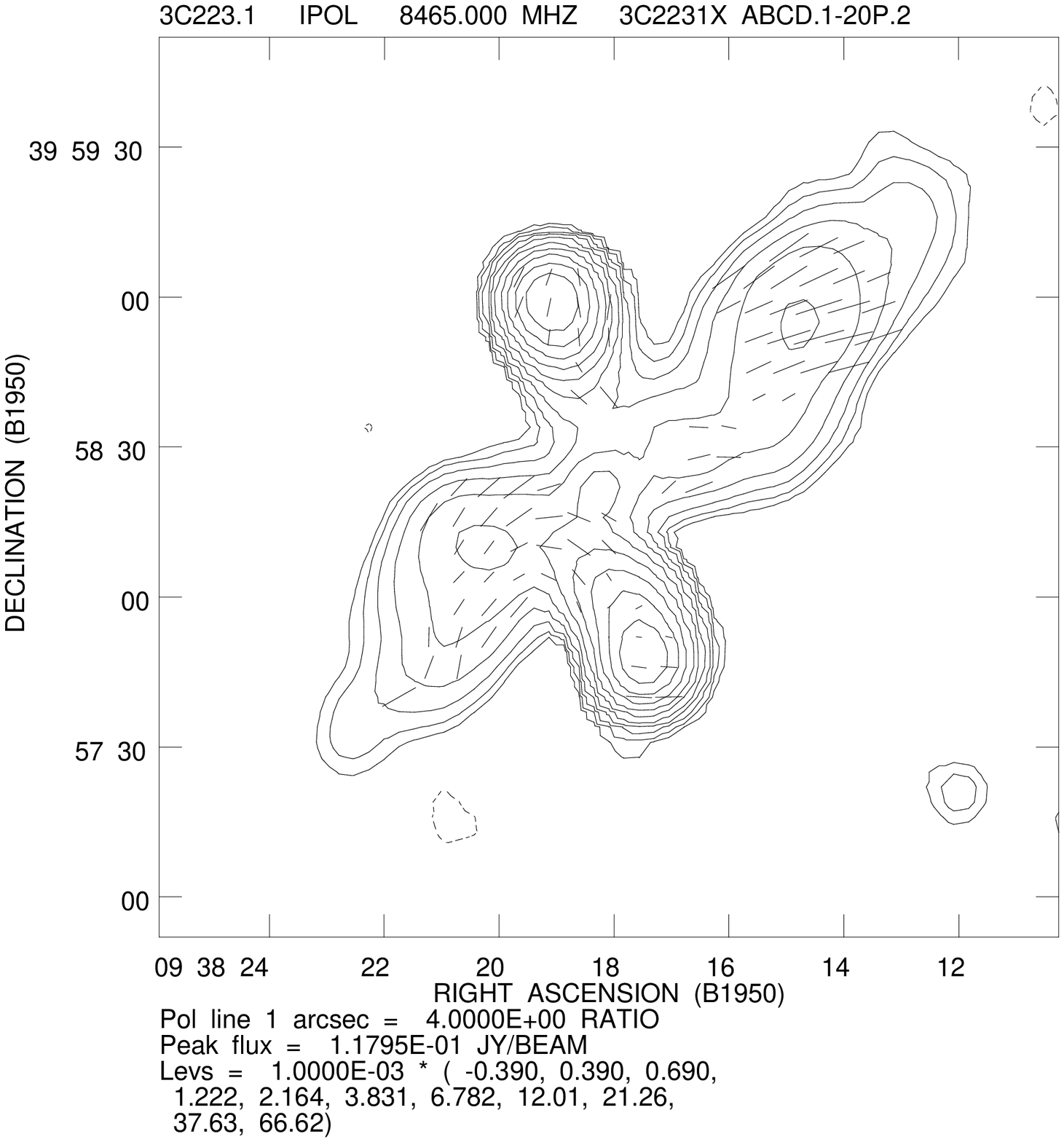,width=7.5cm,angle=-0,clip=}&
(d)\raisebox{6.5cm}{
\epsfig{file=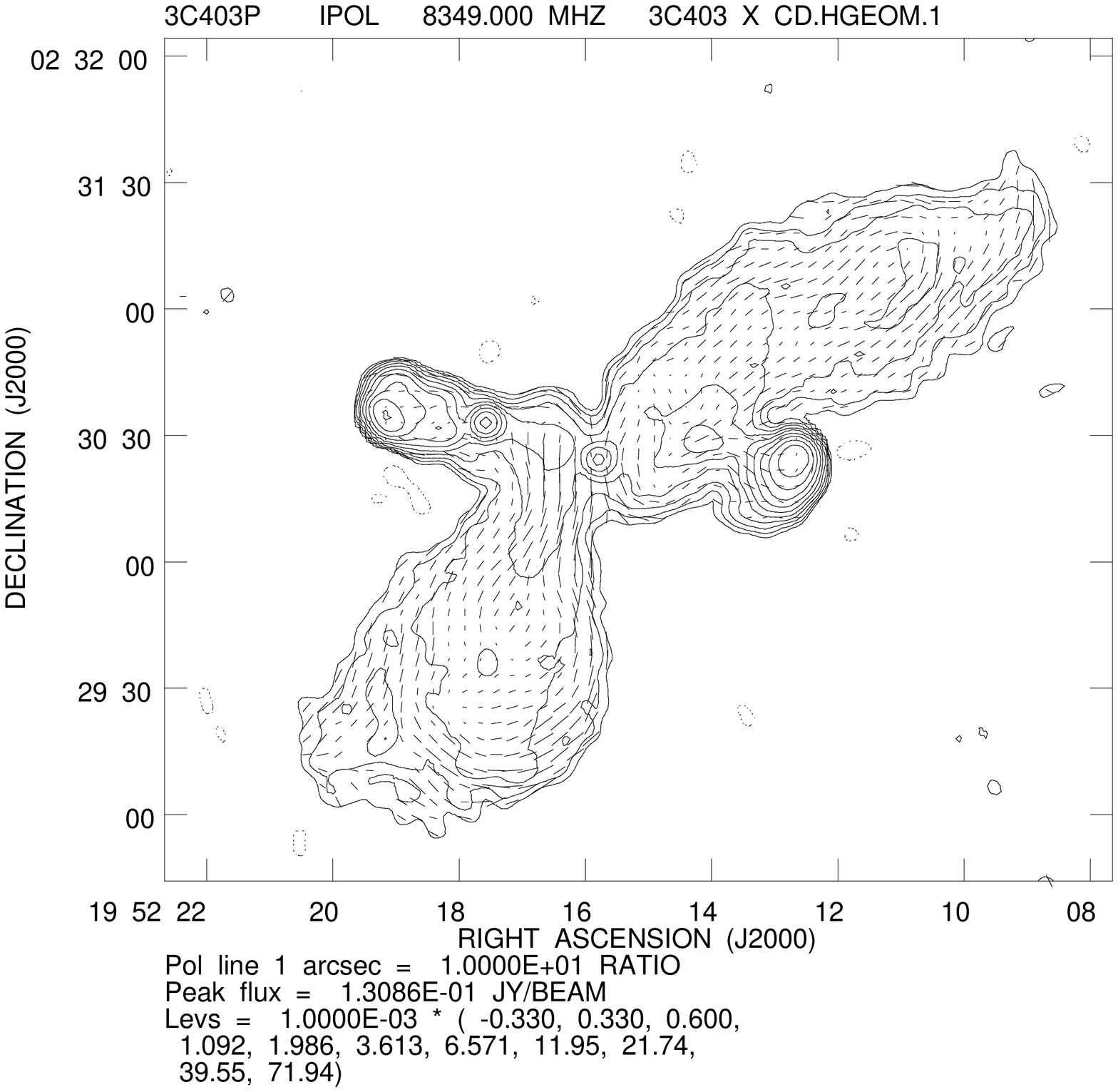,height=7.5cm,angle=-90,clip=}}\\
(e)\epsfig{file=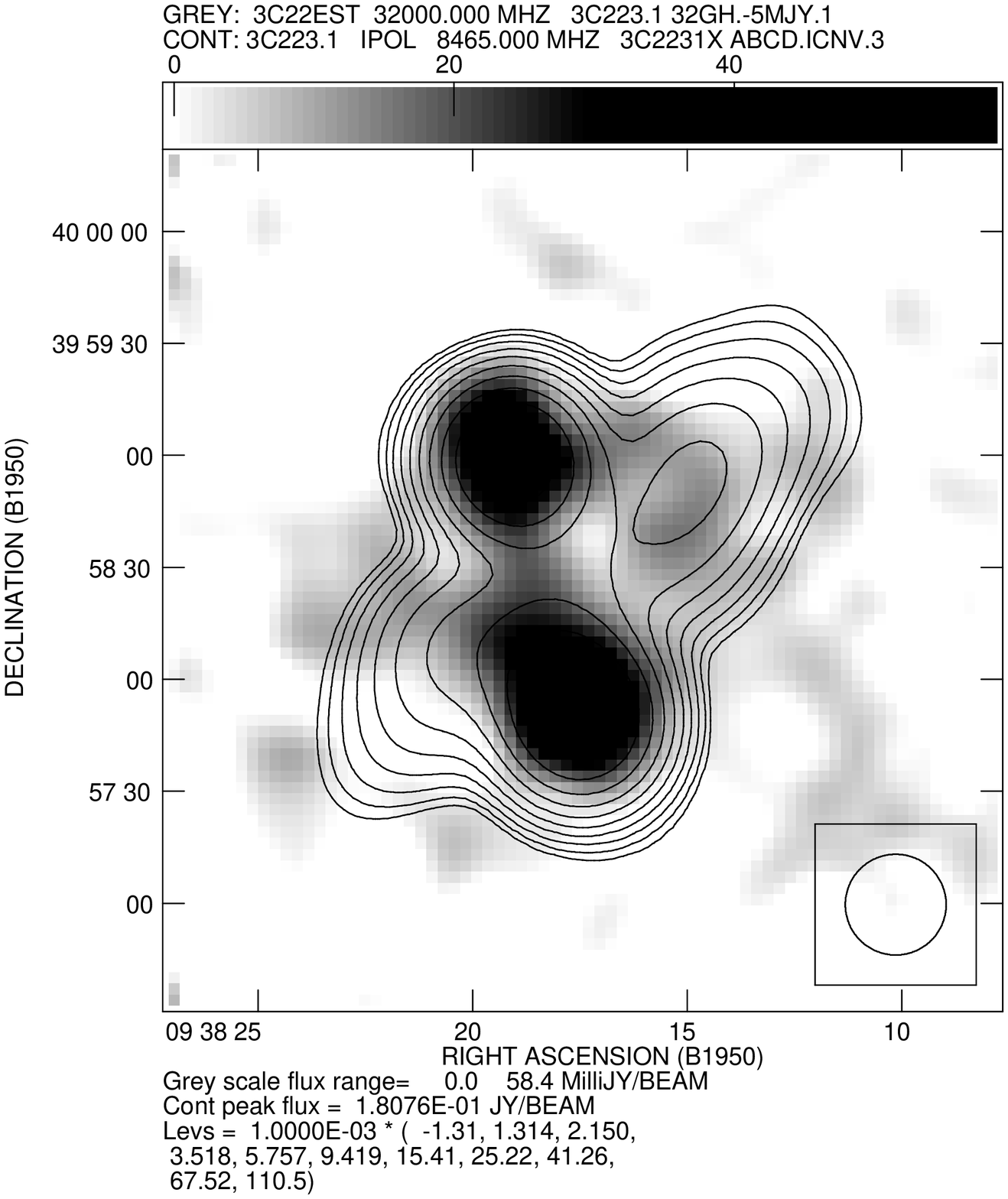,width=7.5cm,clip=} &
(f)\epsfig{file=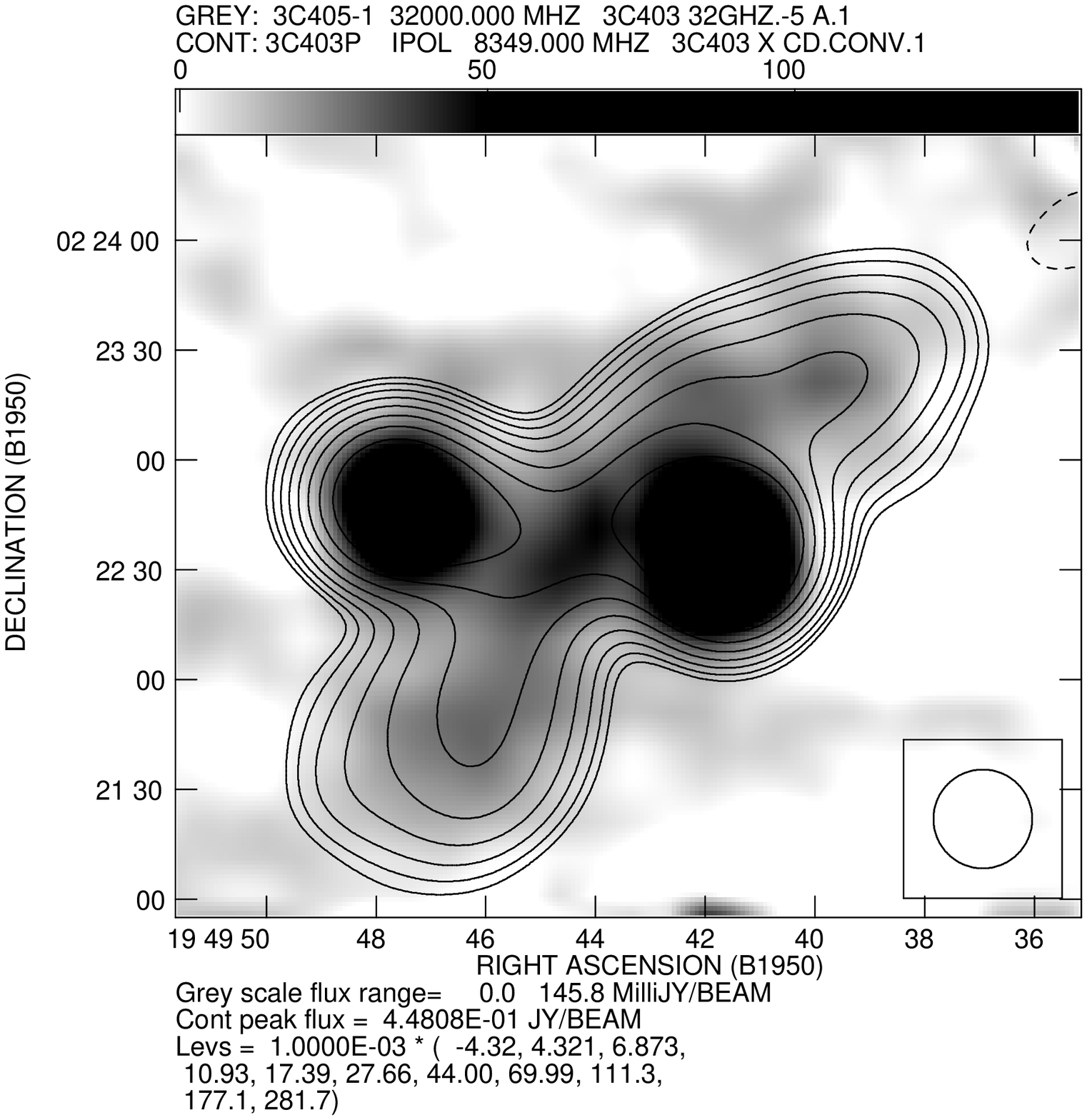,width=8cm,clip=}\\
\end{tabular}

\caption{ {\bf (a)} \& {\bf (b)} total intensity (contours) and
fractional polarization (vectors) at 1.4GHz of 3C\,223.1 \& 3C\,403
respectively. Ten contour levels from 3$\sigma$ to peak flux
density. 3C\,223.1: lowest contours $\pm$ 1.092\,mJy/beam, logarithmic
spacing factor 1.81, vector length 1\,arcsec represents 5\%
polarization. 3C\,403 contours from $\pm$ 0.473\,mJy/beam, spacing
factor 2.02, 1\,arcsec represents 10\% polarization.  {\bf (c)} \&
{\bf (d)} as for (a) \& (b) but at 8.4GHz. 3C\,223.1 contours
$\pm$0.39\,mJ/beam, factor 1.77, 1\,arcsec vector is 4\%
polarisation. 3C\,403 contours $\pm$ 0.33\,mJy/beam, factor
1.82. 1\,arcsec is 10\% polarization. {\bf (e)} \& {\bf (f)} total
intensity at 8\,GHz (contours) and at 32\,GHz (greyscale in mJy/beam).
VLA 8GHz data has been convolved down to Effelberg beam
(27\,arcsec). 3C\,223.1 contours $\pm$1.314\,mJy/beam, factor
1.64. 3C\,403 contours $\pm$ 4.321\,mJy/beam factor 1.59.}
\label{fig:maps}
\end{figure*}

\begin{figure}
\epsfig{file=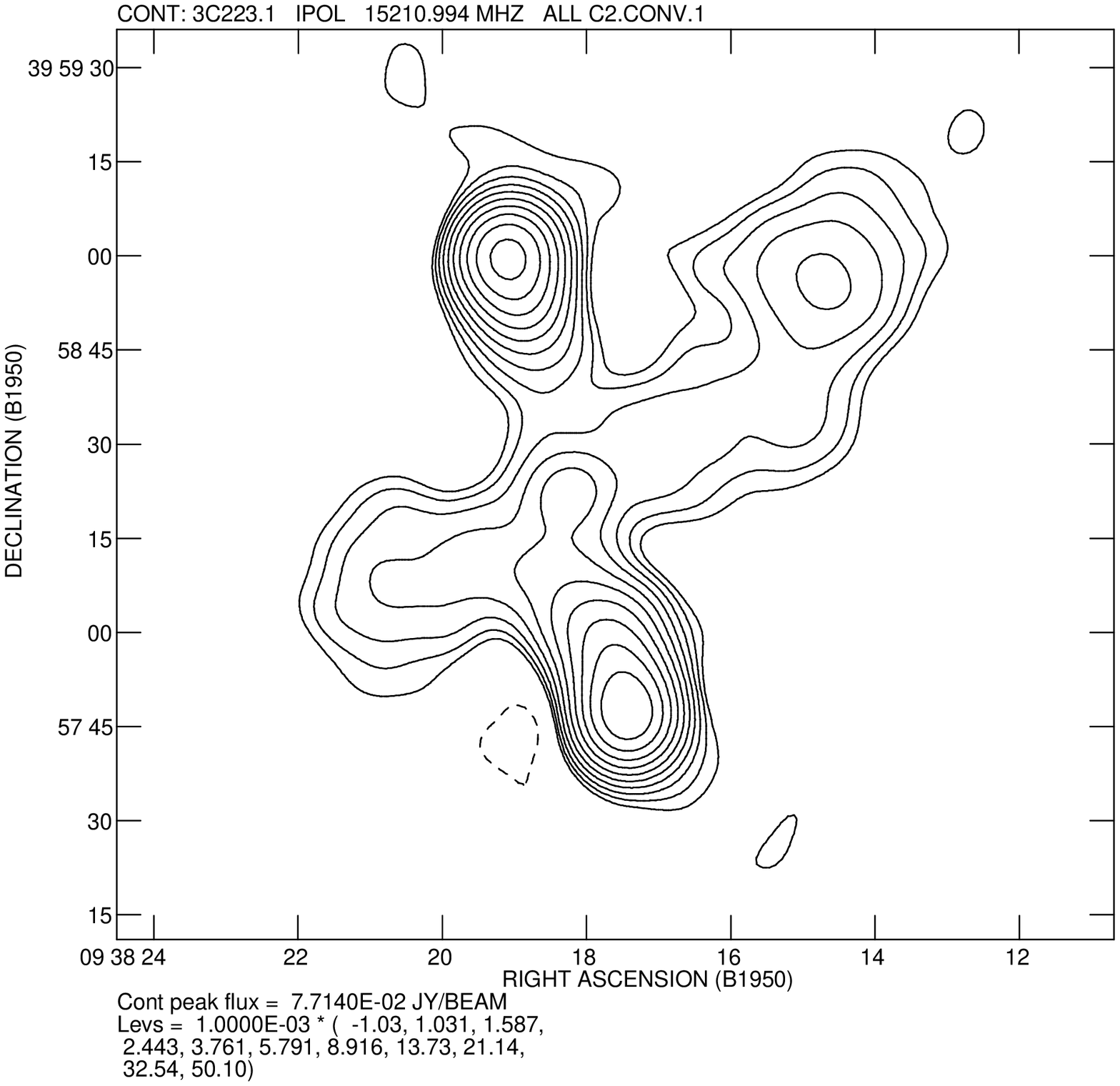,height=7cm, angle=-90,clip=}
\caption{Ryle Telescope observations of 3C\,223.1 at 15.2GHz. Lowest
contour is at $3\sigma$; the 10 contours are equally
  spaced in logarithmic intervals to peak flux density (77\, mJy/beam).
  (Contours: -1.03, 1.03, 1.59, 2.44, 3.76, 5.79, 8.92, 13.73, 21.14,
  32.54, 50.10 mJy/beam)}
\label{fig:3C223.1RT}
\end{figure}

Figs~\ref{fig:maps} (a) --(d) show total intensity images at 1.4 and
8\,GHz, with the apparent magnetic field vector superimposed
(these are rotated by exactly 90$^\circ$ from the
E-vector directions and  uncorrected for any Faraday rotation; see section
\ref{sect:RM}). Figs~\ref{fig:maps} (e) \& (f) show the 32\,GHz total
intensity (greyscale) overlaid on 8.4\,GHz observations convolved to
the same resolution. Fig~\ref{fig:3C223.1RT} shows the 15\,GHz total
intensity image.

Both sources show a high degree of polarization (15--30\%)
in the wings, and an apparent magnetic field structure parallel to the edge of
the source, and along the length of the wings (see also
sect. 3.3). Higher resolution images (B92) showed similar wrapping
around the source edges, as well as complex internal structure. The
high degree of polarization and wrapping of field lines around the
edges of the source can be explained as compression of tangled fields
at the surface of the lobes \cite{lai80a}, and is typical FRII-type
behaviour (e.g. Burch 1979; Miller 1985).\nocite{bur79a,mil85a,per96}

{\bf 3C\,223.1:} This source was imaged at lower resolution than
3C\,403, because it is fainter, and we wished to have high S/N ratio
in the low surface brightness wings. The lower resolution renders the
jet (B92) invisible, and the hotspots comprise a large fraction of the
area and flux density of SW \& NE active lobes.  It is possible that
deeper observations would show that the wings extend further than is
apparent from Figs 1a and c.  The magnetic field vectors in the SW
lobe swing smoothly into the SE wing, aligning themselves with the
ridge of emission that also connects the two components. The northern
components are not as closely connected as the southern components, as
a trough in total intensity separates the lobe and wing. However, the
magnetic field connects the NE lobe and NW wing in the same manner as
the southern components. In these images the the magnetic field
appears to be directed along the length of the wings, even towards the
tips, although here deeper imaging may reveal a more typical `wrap-around'
behaviour.

{\bf 3C\,403:} In the 1.4 and 8.4\,GHz images, which have higher
resolution than those of 3C\,223.1, the E \& W hotspots are clearly
visible, as is the jet-knot in the eastern lobe.  The eastern wing and
lobe are connected by a region of higher surface brightness than the
rest of the SE wing, and by the magnetic field lines which run from
the base of the NE lobe into the SE wing. The pinching-in of the lobes
near the core also appears to separate the eastern and western
components, with the field lines running between the two indentations
and joining smoothly into the fields in the wings and lobes. From the
sharpness of the decline in surface brightness at the edges of the
visible lobes and wings, we deduce that we have not missed any part of
the source which fades into the noise.  The field lines in the wings
of 3C\,403 show loops almost the size of the wings, but not extending
right to the tips. The fractional polarization around the western edge
of the SE wing of 3C\,403 reaches $>$50\% and drops to $\sim$5\% in
the centre of the wing.\\

\subsection{Radio spectra}
\label{sec:spect}

\subsubsection{Spectral index distribution}

Of all the sources in the sample of low-redshift radio galaxies
observed by Dennett-Thorpe {\it et al.} \shortcite{jdt99}, these two
sources have both the lowest minimum spectral index $\alpha_{min}$
(i.e. flattest spectra) and
the least spectral gradient anywhere in the source.

The flux densities plotted in fig.~3 are calculated using images shown
in fig.~1.  For the 1.4\,GHz, 5\,GHz and 8\,GHz images the flux is integrated
over the region above their 3$\sigma$ contour. This is
indistinguishable from results obtained by integrating over the
3$\sigma$ contour at the lower frequency. Because of the poor signal
to noise on the 32\,GHz maps, where the 3$\sigma$ contour does not
meaningfully trace any winged structure, we integrated over the area
enclosed by the 3$\sigma$ contour on the 8\,GHz map convolved to the
same resolution (i.e. the contours on figs 1e \& f).  The flux
densities have been calculated for the entire source, and for regions
of the source separately (i.e. each active lobe and wing) by dividing
the source crudely into these regions. There has been no attempt to
remove the core flux density because this contributes a tiny fraction
of the source flux density at 8\,GHz (0.9\% for 3C\,223.1 \& 0.2\% for
3C\,403; Black 1992).

The spectra in the low surface brightness regions are very sensitive
to changes in the zero-level. For the 32\,GHz single-dish
observations, the error is potentially serious, and a correction to
the mean level was made (see Section~\ref{sec:radobs}). In order to
take account of uncertainties in the correction, we have added an
additional rms error of 5\,mJy/beam in quadrature to the estimate
derived from the images.  At all frequencies, a further 3\% error was
added, again in quadrature, to take account of calibration offsets.

The lengths of the minimum baselines indicate that the largest
structure is sampled at all frequencies, and there are no
indications---such as negative bowls---of missing flux density. The
total flux densities from our observations agree well with values
interpolated from measurements in the literature. The inner \uv plane
is least densely covered in the 8.4\,GHz observations, and we note in
connection with this that a lack of baselines adequately sampling the
source structure will tend to decrease the detected flux at this
frequency. In this case this will falsely {\it steepen} the observed
spectrum, although the use of dual-pointing mosaiced observations
helps to mitigate the effect.

We have obtained the best power-law fit S$\propto \nu^{-\alpha \pm
\sigma_\alpha}$. The results are presented in the first three columns
of table \ref{tab:fit-alf}, including a goodness of fit parameter Q (the
probability that the $\chi^2$ should exceed the calculated value by
chance). If either the 15\,GHz or the 32\,GHz data are omitted, the
results are unchanged to within the errors. The results without
application of offsets at 32\,GHz are also within the quoted errors.

The results for both sources show good fits can be obtained with
power-law spectra all the way to 32\,GHz (with the possible exception of
the northern lobe of 3C\,223.1, where a low 15\,GHz point results in a
poor fit). The fitted spectra have 0.66$< \alpha <$ 0.80 for
all regions of both sources.  3C\,223.1 shows marginal evidence for
steeper spectra in the active lobes than in the wings ($\alpha_{lobe}
- \alpha_{wing} \sim$ 0.08). This result does not depend on the lower
quality 15\,GHz and 32\,GHz observations: the same is true when the
fit is to the 1.4, 5 and 8\,GHz observations alone. This might occur
if flux was missing at lower frequencies due to poorer \uv
coverage, although there is no evidence for this. With the exception of
the western wing of 3C\,223.1, it is also noted that all parts of each
source can be fitted with the same power law.

In no region of either source can we exclude the hypothesis that the spectrum
is a power law over our entire frequency range.  Indeed, there is no evidence
for any steepening of the spectrum in 3C\,403.  In 3C\,223.1, however, we cannot
exclude a break frequency as low as 16\,GHz in the E wing, with the
assumptions given in the following section.

\begin{table}
\caption{Best fit spectra to sources and source regions}
\begin{tabular}{llllll}
\hline
\hline
&\multicolumn{3}{c}{power law fit} &\multicolumn{2}{c}{JP fit}\\
\hline
region&$\alpha$&$\sigma_\alpha$&Q &$\nu_b$(GHz)  & max age (Myr)\\
\hline
3C\,223.1\\
\hline
total& 0.75  & 0.02 & 0.487  &   &\\
N lobe& 0.75   & 0.02  & 0.032 & &\\
W wing& 0.66   & 0.03  & 0.870 & $>$50 &16.6\\
S lobe& 0.77   & 0.02  & 0.254 & &\\
E wing& 0.70   & 0.03  & 0.130 & 12.0&33.9\\

\hline
3C\,403\\
\hline
total&  0.78 &   0.02&   0.313 & &\\
E lobe&0.78 &  0.02 &  0.592 &   &\\
SE wing&0.80 &  0.03 &  0.095 & $>$70 &16.1\\
W lobe&0.77 &  0.02 &  0.451 &   &\\
NW wing&0.77 &  0.03 &  0.275 & $>$60 &17.3\\
\hline
\end{tabular}
\label{tab:fit-alf}
\end{table}

\begin{figure*}
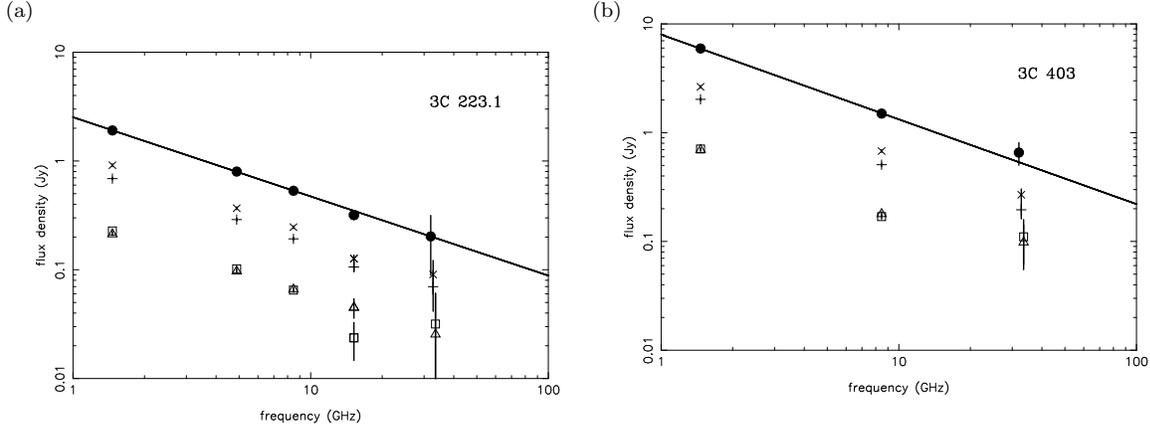

\begin{tabular}{cc}
(a)\epsfig{file=figures/3c2231-S.ps,height=7cm,angle=-90}&
(b)\epsfig{file=figures/3c403-S.ps,height=7cm,angle=-90}\\

\end{tabular}
\caption{Flux densities for (a) 3C\,223.1 and (b) 3C\,403  for the total source (filled circle), the active lobes
  (crosses), and the wings (open symbols). 
The line is best power-law fit for the total flux density ($\alpha$=0.75 and
0.78 respectively).}
\label{fig:fit-alf}
\end{figure*}

\subsubsection{Particle ages}
\label{sec:ages}

Our aim is to set an upper bound to the time since particle
acceleration, using the fact that inverse Compton and synchrotron
losses are comparable in the lobes of 3C\,223.1 and 3C\,403.  We
assume that the initial electron energy spectrum is a power-law
extending to infinite energy, that the electrons lose energy only by
synchrotron and inverse Compton losses and that the magnetic field
remains constant. The age we calculate is then the age since injection
of this electron population. A larger particle age would require that
the magnetic field had been {\it weaker} in the past; in the more
likely event that the magnetic field was stronger in the past, the
true age would be even lower than these limits.  Other complicating
factors, such as continuing input of energetic particles also require
the true ages to be {\em lower} than those we derive here.

\begin{figure}
\epsfig{file=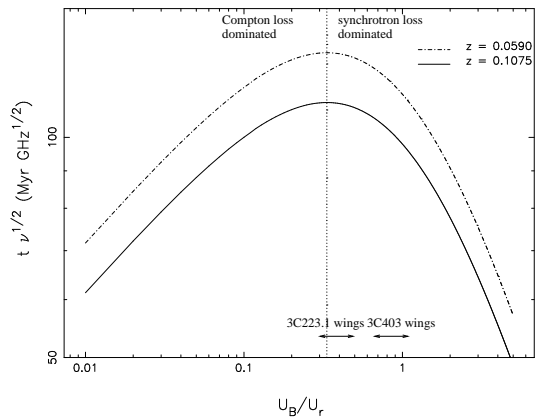,height=7cm,angle=-90}
\caption{Particle ages as a function of the lobe magnetic field and
CMB radiation energy density ratio. The
regimes occupied by 3C\,223.1 and 3C\,403 for equipartition magnetic
fields are indicated. The abscissa is energy ratio and the ordinate a
scaled age for a given {\it observed} break frequency.}
\label{fig:max-ages}
\end{figure}

We regard it as most likely that synchrotron losses occur with pitch-angle 
isotropization (Jaffe \& Perola, 1973). In this case it is straightforward 
to include inverse Compton scattering off the microwave background radiation,
which is independent of pitch angle.  The expression for the break frequency, 
$\nu_b$ is (Pacholczyk, 1970):
$$\nu_b = \frac{(9/4)c_7}{t^2} \frac{B}{(B^2 + B_{CMB}^2)^2}$$ 
where $c_7$ is Pacholczyk's constant = 1.12~10$^3$~nT$^3$\,Myr$^2$\,GHz,
and the equivalent magnetic field due to the microwave background,
$B_{CMB} = 0.32(1+z)^2$nT. Ages as a function of the ratio of
magnetic to CMB radiation energy density are shown in
fig.~\ref{fig:max-ages}.
The break frequency has a maximum value $\nu_b~=~2.54~10^4(1+z)^{-6}t^{-2} $ where
the magnetic field is $B~=~B_{CMB}/\surd{3}$.  

We also fit JP spectra to the wings, using an injection spectrum
$\gamma_{inj}=2.0$ (Table~\ref{tab:fit-alf}, column 4). 
Injection spectra steeper than this will decrease the calculated age.
The spectrum of the source need not be represented by a pure JP
spectrum, but, crucially, there will be an exponential cut--off in the
spectrum with a break frequency corresponding to $\nu_b^{\rm max}$, as
no process can flatten the spectrum at higher frequencies. A range of
magnetic field strengths and/or the absence of electron pitch--angle
scattering can both be considered as mixing in other (effective)
magnetic field contributions. This will steepen the spectrum below
this break, but never flatten the high-frequency tail. By fitting a JP
spectrum we cannot overestimate the break frequency corresponding to
the exponential cut-off $\nu_b^{\rm max}$ and thus our ages are true
upper limits.

If we fit a JP spectrum, we can therefore derive an upper limit to the 
time since particle acceleration which is independent of the magnetic field 
strength (and therefore valid also for a spectrum of field strengths).
The maximum ages we derive for the two sources (column 6
Table~\ref{tab:fit-alf}) are 33.9\,Myr for 3C\,223.1 and 17.3\,Myr for 3C\,403.


\subsection{Rotation measures and Depolarization}
\label{sect:RM}

The wavelengths at half-maximum polarization, $\lambda_{1/2}$,
tabulated for 3C\,223.1 and 3C\,403 by Tabara \& Inoue
\shortcite{tab80} are $>$21cm \& 17cm respectively, suggesting that we
might expect substantial depolarization at 1.4GHz in 3C\,403. This is
not seen.  We compared `quasi-single dish' fractional polarizations
($m = ((\Sigma Q)^2+(\Sigma U)^2)^{1/2}/\Sigma I$) with those
tabulated in Tabara \& Inoue, for all the sources from Dennett-Thorpe
\etal (1999), including the two in the present paper. This revealed two sources
of 20\,cm polarization data used by Tabara \& Inoue which had
consistently low fractional polarizations.
(These are Seielstad \& Weiler \shortcite{sei69} and Bologna, McClain
\& Sloanaker \shortcite{bol69}.)
Our results and all other single dish measurements agree to within the
errors, and imply that $\lambda_{1/2} > $20cm for both sources.

From these observations we have also calculated `quasi-single dish'
polarization position angles ($\chi = 0.5$arctan$(\Sigma U/\Sigma Q$)) and
the changes in these values between 1.4 \& 8\,GHz. Images of the
position angle changes were also produced (for points where the 1.4GHz
total intensity image was above 3$\sigma$). Both these measures give
good agreement with the rotation measures calculated in
Simard-Normandin, Kronberg \& Button \shortcite{sim81a}: 3C\,223.1 has
RM$=+3\pm4$\,rad\,m$^{-2}$, 3C\,403 has RM$=-36\pm1$\,rad\,m$^{-2}$. 

\begin{figure}
(a)3C\,223.1
\epsfig{file=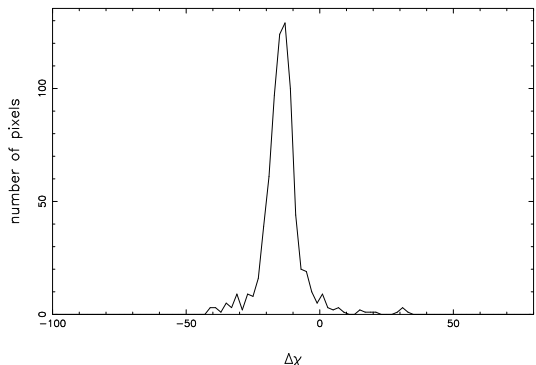,height=7cm,angle=-90, clip=}

(b)3C\,403
\epsfig{file=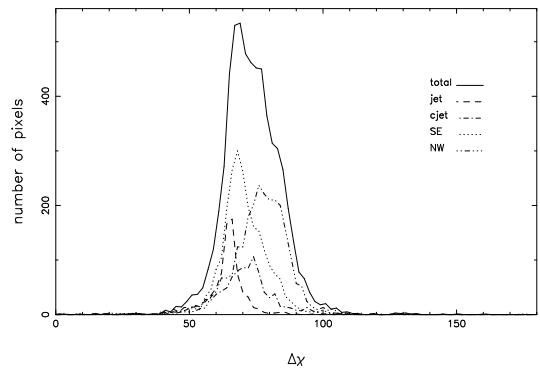,height=7cm,angle=-90,clip=}
\caption{Change of polarization position angle in degrees between 1.4 \& 8GHz,
  everywhere on source (I $> 3 \sigma_I$; 8.4$-$1.4\,GHz). The narrow peak of 3C\,223.1
  indicates little change across the source. The different components of
  3C\,403 are also shown in (b). There is a smooth change in rotation
  measure across the source, but little structure on scales similar to
  the beam, as for 3C\,223.1.}
\label{fig:dp}
\end{figure}

An investigation of the position angle variations between 1.4 and
8.4\,GHz show little change or variation across the source in
3C\,223.1.  In 3C\,403, there is gradient across the source (see
fig.~\ref{fig:dp}), whose magnitude and scale is consistent with the
effect of a foreground screen in our Galaxy
\cite{sim84,lea87}. Also, it is noted that 3C\,403 is seen
through the Scutum Arm of the Galaxy, so a negative contribution to
the rotation measure from this is expected \cite{val88b}.

\begin{figure}
\epsfig{figure=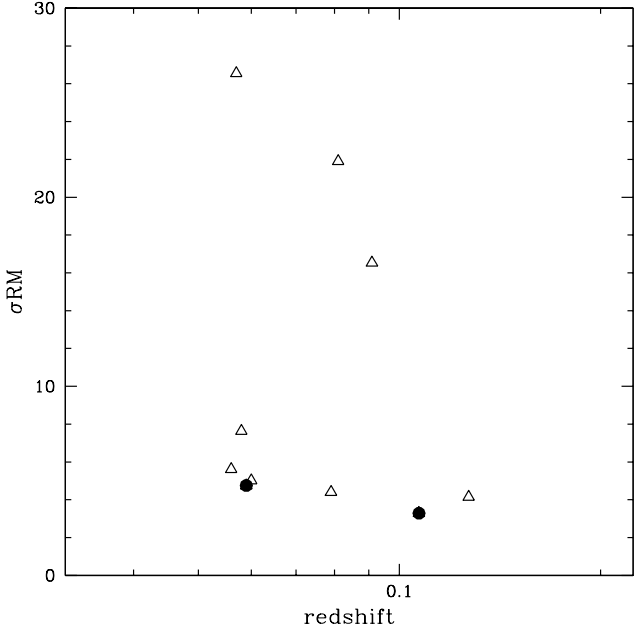,height=9cm,angle=0}
\caption{The standard deviation of the changes in rotation measure
(i.e. as fig.~\ref{fig:dp}) in a sample of sources at similar redshift and power, observed in
a similar manner.}
\label{fig:dchi-z} 
\end{figure}

In order to assess the likely environment of the two sources, we
compare the standard deviation of the polarization position angle
changes to those found using similar data for a sample of 3C sources
in the same P--z range \cite{jdt99}. The results are presented in
fig.~\ref{fig:dchi-z}. There are two sources with considerably higher
variations of polarization position angle: 3C\,388 and 3C\,452.
3C\,452 is in a region of sky in which extraglactic sources have a
large excess of negative rotation measures \cite{sim80}. It is likely
that the (de)polarization structure seen in 3C\,452 is due to this,
probably Galactic, feature \cite{jdt_thesis}. 3C\,388 is the only
source of this sample known to lie in a rich cluster
\cite{fab84}. 3C\,405 (Cygnus\,A), a member of the sample for which we
did not have comparable data, would also have shown a high variation
of PA \cite{dre87}.  From various studies of clustering in the
literature all the other sources appear to be isolated, or at best lie
in open groups and poor clusters. Thus, the low RMs and low
depolarising frequency are consistent with both sources being in
relatively gas poor environments, no richer than a poor cluster.

\section{Comparison with other sources}
\label{sec:comp}

Leahy \& Parma \shortcite{lea92} showed that winged sources (defined by the
condition that the wings are at least 80\% of the length of the active
lobes) make up $\approx$7\% of the radio galaxy population with
luminosities between 3 $\times$ 10$^{24}$ and 3 $\times$ 10$^{26}$
WHz$^{-1}$ at 1.4\,GHz.  They identified 11 winged sources, of which 7
have clear FRII structure, 2 are borderline cases more obviously
related to FRIIs, 1 is too distorted to classify and only 1 (NGC\,326;
see below) is an unambiguous FRI.  The phenomenon is therefore
primarily associated with low-luminosity FRII sources.

In this section we contrast the properties of a number of well-studied
sources showing evidence for change of jet direction or episodic
activity, with those of 3C\,223.1 and 3C\,403.\\

\noindent{\bf B2\,0828+32}

The low luminosity radio galaxy B2\,0828+32 (S$_{1.4GHz}$~=~2.1Jy; z =
0.0507) is morphologically very similar to the sources discussed here
and has been the subject of a number of studies.  Its active lobes
project 320\,arcsec (430~kpc) and its wings at least twice this
distance.  In addition to the overall morphology, the magnetic field
structure is very similar to those of the sources presented in this
paper. In particular, it exhibits high fractional polarization in
the wings (up to 40\%), alignment of the magnetic field along the
wings, and little depolarization between 0.6 and
10\,GHz\cite{par85,gre92,mac94}.

The active lobes have a fairly flat spectrum which steepens in the
wings: $\alpha^{1.4GHz}_{0.6GHz}$(lobes) $\sim 0.7$,
$\alpha^{1.4GHz}_{0.6GHz}$(wings) $\sim 1.2$ \cite{par85,gre92}.  This
is confirmed by other data, including 10.6~GHz Effelsberg observations
\cite{mac94,kle95}.  From WSRT 0.6GHz \& Effelsberg 10.6\,GHz data,
Klein \etal \shortcite{kle95}, find very little change in spectral
index over the active lobes, but a significant steepening of the
spectrum towards the tips of the wings (except for a flat-spectrum
feature in the northern wing). On this basis the authors argue in
favour of formation in terms of a slow conical precession.

The break frequency in B2\,0828+32 is lower by a factor of ten or more
than in the sources studied here\cite{kle95}, implying a greater
maximum time since particle acceleration (the redshift is comparable
to that of 3C\,403).  Evidence -- albeit weak -- for a galaxy-galaxy
interaction within the last 10$^8$ years was found by Ulrich \&
R{\"o}nnbach \shortcite{ulr96}.\\


\noindent{\bf 3C\,293}

3C\,293 is a borderline FRI/FRII structure with an extended $\sim$ 200
 \,kpc source \cite{bri81} around a smaller $\sim$ 2\,kpc misaligned
 double structure within the steep spectrum core \cite{aku96}.  Its
 structure could therefore be interpreted as an extreme case of a
 winged source with very short active lobes.  In contrast to
 3C\,223.1, 3C\,403 and B2\,0828+28, the host galaxy of 3C\,293
 (z=0.0452) shows obvious signs of interaction \cite{mar99,eva99}: it
 has a tidal tail, a nearby companion and copious amounts of dust and
 CO.  Although it has been suggested that the misaligned structure
 might be produced by pressure gradients or collisions with gas clouds
 \cite{bri81,bre84}, this interpretation encounters problems if the
 jet is relativistic \cite{eva99} and the hypothesis of change of jet
 direction again seems more likely.

It is intriguing that the source with the shortest ratio of active
lobe length to wing length is the only one which shows clear evidence
for a recent merger.\\


\noindent{\bf Double--double giant radio sources}

Recently Schoenmakers et al \shortcite{sch99a} have shown that a
number of giant radio galaxies have a complex structure which includes
a second, inner pair of radio lobes very similar to the outer pair. In
these sources the angle of the new jet axis is with 10$^\circ$ of the
old jet axis. In B\,1834+620, the presence of a hotspot in one of the
outer lobes allows an upper limit of a few Myr to be placed on the
time for which the central activity was `turned off' before entering a
new phase of activity \cite{sch99b,lar99}.

These are possibly close relatives of the winged sources, because they
too display multi-phase activity, which clearly has a nuclear
origin. However, the unchanged alignment of the source axis between
phases in the double-double giants known to date may indicate an
important difference.\\

\noindent{\bf Wings in FRI sources?}

NGC\,326 \cite{eke78,wor95}, a nearby FRI radio galaxy, is often
compared with the winged sources because its morphology implies a
change of direction of the jet axis.  In this case, there are no
excresences starting close to the core, but rather a change of
direction in the radio plasma where the jets terminate. Although 
this is perhaps not very clear on the northern side of the
source, it certainly appears to be the case on the south.  For this
reason has been called a `Z-shaped source', and is morphologically
distinct from the sources considered here. Leahy \& Parma
\shortcite{lea92} note that Z- or S-shaped distortions are commonplace
in FRI sources. It could be that the mechanism of jet realignment is
the same in both FR classes, but that differing mechanisms of lobe
formation lead to the two observed morphologies.\\

We conclude from this comparison that the phenomenon of wings appears
to be restricted to low-power FRII sources.  It may be that 3C\,293,
which is undergoing a merger, may be an extreme example. Evidence for
multiple phases of activity without significant changes of jet
direction is provided by the double-double sources. In contrast,
changes of orientation do appear to occur in FRI sources, but give
rise to S-shaped distortions rather than wings.


\section{Formation models: Constraints from the radio data}
\label{sec:form}

\begin{figure*}
\psfig{file=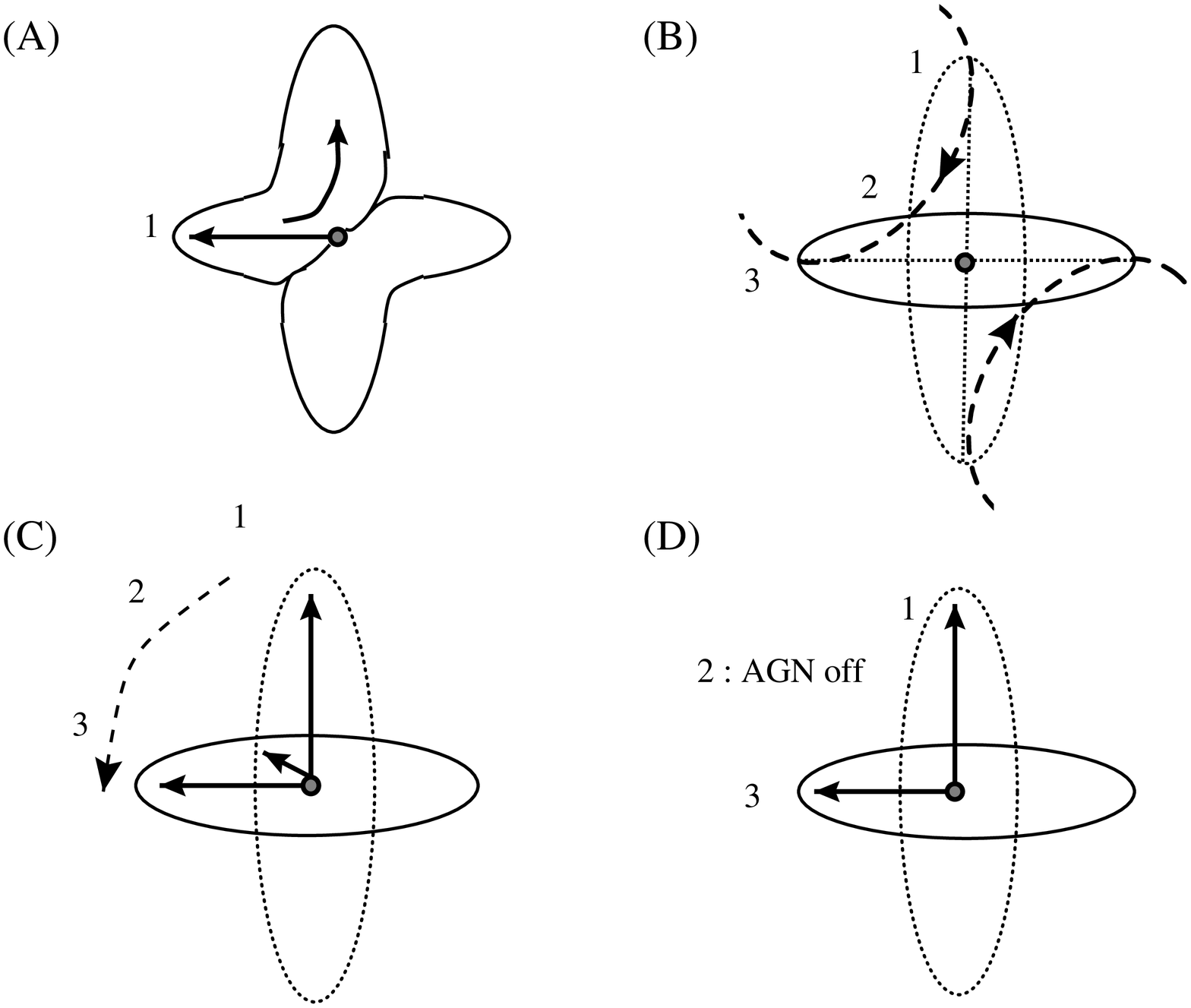,width=15cm}
\caption{Sketch of formation models. (A) Backflow from active
lobes. (B) Slow `conical' precession of jet axis. In order to see the X-shaped
morphology the surface of the precession cone has to pass close to the
line of sight at point 2. (C) A reorientation of the jet axis, during
which the AGN continues jet production. (D) A reorientation of the jet
axis between two distinct phases of AGN activity.}
\label{fig:models}  
\end{figure*}

In fig.~\ref{fig:models} we present a sketch of the four formation
scenarios discussed here. These are:
\begin{description}
\item [A] Backflow from the active lobes into the wings.
\item [B] Slow, conical precession of the jet axis.
\item [C] Reorientation of the jet axis during which the flow continues.
\item [D] As model C, but with the jet turned off or at greatly reduced
power during the change of direction.
\end{description}

If spectral gradients had been detected in our sources, they could
distinguish between some of these models.  In the sudden reorientation
model (C) or multiple burst model (D), with no particle reacceleration
or backflow into the relic lobes, the oldest relativistic electrons
would be near the centre, whereas the pure backflow model (A) would
have the oldest particles at the tip of the wings. Leahy \& Williams
\shortcite{lea84} also suggested a hybrid model in which wings are formed by
backflow into cavities of `relic lobes'. The addition of backflow might
leave a spectral signature, but would change the picture from the
simplest case described above. The more surprising result found here,
that no spectral curvature appears up to 32 GHz, will be used below to
constrain the models in other ways.

Attempts to use the polarization characteristics of the sources to
constrain their formation are frustrated by the dissimilarity of the
sources. While all wings show remarkably high percentage polarization,
there is considerable difference in the distribution and
alignment. (This difference remains when 3C\,403 is convolved down to
the same linear resolution as is used for 3C\,223.1.) In 3C\,223.1 the
projected magnetic field runs parallel to the wings, and is maximal in
the centre. In 3C\,403, however, the field `wraps--around' the edges
of the wings, and falls in the centre of the wings.  The resulting
field configuration is likely to have a complex dependence on the
initial conditions, backflow and expansion, which cause compression,
expansion and shear of the field.  What is not clear is why one should
dominate in one source, the other in another.

\subsection{Backflow and buoyancy}
\label{sect:buoy}

The idea that excrescences near the host galaxy are overshoots of the
backflow of radio-emitting plasma along the active lobes (Model A)
(Leahy \& Williams 1984) is an attractive one for relatively small
excrescences such as are found in 3C\,382 and 3C\,192 \cite{bau88}.
It becomes untenable for the wings of sources such as 3C\,223.1 and
3C\,403, where the wings are longer than the active lobes. (The active
lobes are not strongly foreshortened by projection: see Section 1.)
As realised by Leahy \& Parma \shortcite{lea92}, backflow combined
with buoyancy effects \cite{gul73} cannot be solely responsible for
the formation of the wings: buoyant expansion is subsonic, and
therefore cannot push faster than the active lobes, whose expansion is
driven by strong shocks at the hotspots. A very peculiar distribution
of IGM with a strongly non-spherical distribution would be required.

Buoyant forces could potentially cause the expansion of relatively
small initial wings (e.g. produced by backflow). However, wings formed
by such expansion would rapidly fall below the limit of detection; the
flux density of an adiabatically expanding lobe falls at least as fast
as (linear size)$^{-4}$. (The decline is steeper in the case of
downcurved spectra.) Consider a generous estimate of the size of the
maximum feasible backflow wings: that they were originally the size of
the active lobes, and have buoyantly expanded to their present state.
We see that the wings must have been at least $\sim$ 5 times as
luminous as the presently active lobes, and therefore conclude that
they could not have been backflow excresences, but must have been fed
by jets directed along their axes. From this calculation, we also see that
the jets must have injected particle and magnetic energy at 
significantly higher rates into these first lobes, that these were
created in a significantly denser medium than the presently active
equivalents, or (most likely) that they were originally larger in extent
than the presently active lobes.

\subsection{Conical precession}
\label{sect:cone}

The conical precession model (B in fig.~\ref{fig:models}, Parma \etal 1985; 
Mack \etal 1994)\nocite{par85,mac94} requires not
only a fortuitous angle between the precession cone and angle to the
line of sight, but also a happy accident of the positions at which the
source first switched on, and its position now. It therefore seems
unlikely that we can explain the number of such sources seen, or the
lack of other related sources (Leahy \& Parma, 1992).

In detail, the morphologies of these two sources do not seem to fit
this model either. A relatively high surface-brightness structure
linking the wing and lobe would be expected if the morphology arose
from a special projection of a slowly precessing source. With the
exception of the SE region of 3C\,223.1, there is a notable lack of
such a feature.  The total intensity images may indicate connection
between the wings and the lobes, but there is a notable asymmetry this
respect in 3C\,223.1 (a trough in total intensity on the NW connecting
region, and a smooth, bright region in the SE).  Further, the
pinching-in of the wings at the base also argues against any
interpretation in terms of slow motion of the jet axis (both models B
\& C).  Given these arguments, it is more likely that such slow
precession can be used as an explanation for Z-shaped sources such as
NGC326 (Sect.~\ref{sec:comp}), than sources such as 3C\,403 and
3C\,223.1.  \nocite{bla92b,eke78}

\subsection{Speeds of expansion and realignment timescales}
\label{sect:realign}

Evidence from surface brightness distributions and morphologies argues
for sources whose radio-jets have undergone a sudden reorientation in
a small fraction of its lifetime (Models C \& D).  In so far as no
fresh relativistic electrons have been put into the wings (e.g. by
particle reacceleration), the maximum ages obtained in Section
\ref{sec:ages} are firm upper limits.  Using these upper limits, we
derive a lower limit on the average (hotspot advance + expansion)
speed of 0.022$c$ and 0.035$c$ for 3C\,223.1 and 3C\,403 respectively.

Presumably the time taken to develop the present active lobes is
similar to the time taken to develop the wings when they were the
active lobes, so a generous upper limit for the time taken by each
is 20\,Myr. The distinctness of the wings and the present active lobes
indicates that realignment took place over a time less than 1/4 of
that time, i.e. 5\,Myr.

Alternatively, it is possible the the jets are switched off during
realignment; realignment could then have taken longer. Age estimates
of powerful radio galaxies using different techniques over a wide
range of radio powers \cite{ale84,sch95,par99} agree that expansion
speeds for low luminosity FRIIs are in the range from 0.002 to 0.04c.
Using expansion speeds of 0.04c the estimated off-times are a few
Myr. Whilst a more typical (smaller) expansion speed would decrease
this off-time, we cannot rule out the possibility that the sources
expand unusually fast.  Therefore we cannot set stringent limits to
the realignment time on observational grounds. However, in the next
section we shall see that disc physics indicates a realignment time
much shorter than 5\,Myr, if the black hole and disc parameters are in
the commonly assumed ranges.

\subsection{Summary}

We have argued that the wings seen in 3C\,223.1 and 3C\,403 cannot be
produced by deflected backflow, and that, due to the observed
brightness of these features, buoyant expansion cannot play a large
role in their formation. A conical precession model is excluded on
morphological grounds. The lack of pronounced spectral gradients out
to 32\,GHz, even in the wings implies ages no more than a couple of
tens of Myr, and average advance speeds of $>$2\%$c$.  The jet
axis underwent a major reorientation in both sources, which occurred
on timescales of no more than a few Myr.

\section{Mechanisms for rapid jet reorientation}
\label{sec:mech}

If the jet is aligned with the spin axis of the black hole or the
inner accretion disc, its axis could change as a result of: a merger
with another black hole; interaction between the black hole and a disc
with a different spin axis or instability in the disc.  Observational
evidence for any merger event which could provide either the needed
accreting material, or the second black hole, is briefly examined in
the following section, before comparing with theoretical models for
realignment timescales and brief consideration of possible causes.

\subsection{Evidence against mergers}
\label{sec:merge}

All available evidence suggests that both sources are located in
sparse environments. Rotation measures are small and plausibly
associated with our Galaxy (Section 3.3); there is little hot gas
and the redshifts of galaxies near 3C\,223.1 do
not indicate the presence of a cluster \cite{bur81}. The most obvious
source of the trigger for the rapid realignment or multiple outburst
is a galaxy--galaxy interaction. In such poor environments the chance
of interaction with neighbouring galaxies is substantially decreased.

A search for evidence of recent galaxy--galaxy interactions in both
galaxy hosts, using multicolour optical continuum and emission-line
images from the literature, revealed no particularly impressive signs
of disturbance, companions or multiple nuclei
\cite{bau88,smi89b,smi90,mcc91,kof96,mar99}. Furthermore, long--slit
spectroscopy reveals 3C\,403 to be relatively undisturbed compared
with other radio--loud galaxies\cite{bau90}.  

A large merger event is expected to be still visible, possibly
photometrically, but certainly dynamically (e.g. Balcells \& Quinn
1990). The rotation timescale of the host galaxy is of the order
10$^7$yrs (Barnes \& Hernquist, 1996), whilst a typical merger is only
complete after a few times $10^8$yrs: considerably longer than the
maximum particle ages. 3C\,403 has unusually high ionisation lines
\cite{tad93}, as well as being an IRAS source, but the significance of
these facts is unclear.  The available evidence indicates that an
interaction with a large galaxy was not the immediate cause of any
realignment of the jet axes of these sources, unlike 3C\,293.

\subsection{Black hole realignment with accretion disc}
\label{sec:realign}

It will be convenient in what follows to refer to the axis of the
accretion disc as `vertical'.

A rotating black hole tries to realign material in a misaligned
accretion disc via the differential precession due to Lense-Thirring drag;
viscous forces damp the differential precession, so that a steady state
is reached in which the disc axis is aligned with the black hole axis
out to the warp radius $R_{\rm warp}$ (Bardeen \& Petterson 1975).
Intuitively one expects that the angular momemtum change in the accreting
material near $R_{\rm warp}$ produces an equal and opposite reaction in
the angular momentum of the black hole, but the physics is surprisingly
different. Here it is important to distinguish between the azimuthal
viscous forces in the disc (kinematic viscosity $\nu_1$) and viscous
forces in the vertical direction (kinematic viscosity $\nu_2$).
Calculations showed that both $R_{\rm warp}$ (Papaloizou \& Pringle 1983)
and the time for realigning the black hole (Scheuer \& Feiler 1996)
depend only on $\nu_2$. The reason is that the inward advection of
angular momentum (via $\nu_1$) is rather accurately cancelled by the
outward viscous transport of angular momentum due to $\nu_1$, and the
warp is determined entirely by balancing the Lense-Thirring drag against
the vertical viscous forces in each annulus. The realignment time
is similar to the much earlier estimate by Rees (1978) if $\nu_2 = \nu_1$,
but the different physics has striking results if $\nu_2 \gg \nu_1$: the
realignment time is then much shorter ($\propto \nu_2^{-1/2}$) though
$R_{\rm warp}$ is much smaller ($\propto \nu_2^{-1}$).
\nocite{bar75,bla99,pap83}

There are indeed good reasons to believe that $\nu_2 \gg \nu_1$ in
Keplerian discs (Papaloizou \& Pringle 1983, Kumar \& Pringle 1985).
The radial pressure gradients due to the warp set up radial flows whose
natural period resonates with the period of the applied force (the orbital
period) and therefore reaches large amplitude. The consequent vertical
loading is equivalent to a vertical viscous force characterised by
$\nu_2 = \nu_1 /2\alpha^2$ where $\alpha$ is the Shakura-Sunyaev viscosity
parameter. Combining the above with a disc model,
Natarajan \& Pringle \shortcite{nat98} obtain formulae equivalent to
\nocite{kum85}

\begin{eqnarray}
 t_{\rm align} (yrs) =  3.6\times 10^4
 a^{11/16}\left(\frac{\alpha}{0.01}\right)^{13/8}
 \left(\frac{\epsilon L_E}{L}\right)^{7/8} M_8^{-1/16} \\
 R_{\rm warp}  =  22 
 a^{5/8}\left(\frac{\alpha}{0.01}\right)^{3/4}
 \left(\frac{\epsilon L_E}{L}\right)^{1/4} M_8^{1/8} R_{\rm Schwarzschild}
\end{eqnarray}

\noindent where $a$ is the black hole's angular momentum as a fraction
of the maximum, $\epsilon$ is the efficiency of converting accreted
mass into energy, $L$ is the power output, $L_E$ is the Eddington
luminosity and $M_8$ is the black hole mass in units of $10^8$ solar
masses.

From the minimum energy in the wings in the form of relativistic
electrons and magnetic field and our maximum age estimate of 30\,Myr
we find that in 3C\,223.1 and in 3C\,403 the energy output of the AGN
is at least 0.5\% of $L_E$ for a $10^8 M_\odot$ black hole.  Therefore
this model predicts that realignment can occur in less than 0.5\,Myr for
$a=1, \; \alpha=0.01, \; \epsilon=0.1, \; M_8=1$. The minimum energy
density in the wings is so low that they may well be pressure
confined; in that case the energy estimate should be multiplied by roughly
4/3 to take account of the work done in making room for the wings, and
the realignment time reduced accordingly. These estimates of
realignment times are well within the upper bound of 5\,Myr obtained in
Section \ref{sect:realign}.

All of the above refers essentially to thin radiative accretion discs.
It may be argued that it is therefore irrelevant to these weak radio
galaxies, because they contain ADAFs (in which the incoming matter
falls into the BH without radiating, carrying its heat content with
it) or ADIOs (in which most of the incoming matter flows out again at
higher latitudes) (Blandford \& Begelman 1999 and references
therein). Numerical simulations by Stone et al. (1999), and in papers
by Igumenshchev and collaborators cited there, are qualitatively more
like ADIOs in that most of the matter flowing inward through the disc
flows off the computing grid at high latitudes, though the power law
dependences of density, pressure, etc. with radius are different. For
our present purposes, the essential features that all these models
have in common is that they have very thick non-radiative discs. None
of them give clear guidance on the fraction of infalling matter
accreted, as everything scales in power laws in radius (which result
from the simulations, and are imposed, through self-similarity, in the
analytic models).  Furthermore, they show little promise of going over
to radiative discs at small radii, since (even in the more favourable
models) the increase in density at small radii is insufficient to make
up for the shorter accretion time scale and thus lead to effective
Coulomb transfer of heat to electrons. Nevertheless, observation
forces us to conclude that these radio galaxies have radiative discs,
at least at small radii. The AGN emit X-rays and, although broad-line
emission is not detected, the narrow-line emission is strong,
requiring a hidden source of ionising photons. Therefore we still
adopt the above estimates of realignment times, taking $\epsilon$ to
refer to the efficiency in terms of the mass input to the radiative
part of the disc.  \nocite{sto99}

\subsection{Switching jet direction without black hole realignment}

It is possible that the jet axis is not determined by the alignment of
the black hole itself, as we have assumed until this point.  If
the paths taken by components in well-studied superluminal quasars
such as 3C\,345 \cite{zen95} represent the jet axis rather than some
trajectory within a wider jet, then the axis changes direction on a
timescale of a few years, much shorter than any reasonable time
scale for black hole realignment, and we must conclude that the jet
direction is strongly influenced by the disc. If the jet axis was at
different times controlled by different parts of the disc, or the
inner disc or the jet axis itself suffered instablities a switching of
the jet direction could take place over a short time scale.

Such a situation could arise for example if $\alpha \sim 0.1$ and the
black hole mass is well above $10^9 M_\odot$ (so that $L_E$ is
correspondingly greater): in this case the black hole realignment time
is several times $10^7$ years.  Since quite extreme values of the
parameters are needed to prevent realignment of the black hole on
the timescales of concern here, we regard this possibility as unlikely.

\subsection{The origin of the realignment}

If the change in angular momentum of the accretion disc is the `prime
mover', either infalling matter of a different angular momentum, or a
disc instability could be the cause. An ingested dwarf galaxy may not
have left the observable traces that a larger merger would, and could
have provided sufficient mass and angular momentum to produce the
change in jet axis. It is also possible that the gas from an earlier
merger has now finally reached the very central regions.  A further
possibility is an instability of the accretion disc (e.g radiative
warping instability \cite{pri96}), but we then have to explain why the
instability seems to be suppressed in other radio galaxies which have
stable jet directions and why it occured only once, briefly, in the
winged sources.

A second possibility is that the black hole itself is the `prime
mover'. A coalescence of a binary black hole system would nicely
explain the single major reorientation of the jet axis. If a galaxy
merger occurred $>$ 10$^8$yr ago the gaseous traces of this would no
longer be visible and the central black holes would settle into a
binary orbit \cite{beg80}.  Estimates of the timescale for the
coalescence of two black holes seem uncertain, and range from
$\sim10^7$yrs \cite{ebi91} to longer than the Hubble time
\cite{val89}. The timescale which we have called the `reorientation
timescale' is likely to be only the time between the gravitational
potential of the secondary black hole significantly disturbing that of
the first, and coalescence.

The detection of `wings' only in low power FRII sources may simply be
a selection effect. Synchrotron losses are likely to be much more
rapid because the magnetic fields are likely to be stronger (both
because of their higher radio power and higher redshift). Furthermore,
adiabatic losses would be severe for wings expanding supersonically
into the IGM, while the low-power wings we observe are likely to be
pressure-confined and would not expand significantly. 
We would then expect that wings would be detected in more powerful
sources at very low frequencies. It is much more difficult to devise
selection effects to explain why structures analogous to wings are not
seen in the less powerful FRI sources. 

\section{Conclusions}

We have presented high quality images at 1.4 and 8.4\,GHz of two radio
galaxies with a peculiar `winged' morphology in the radio plasma,
3C\,223.1 and 3C\,403. We have also presented data at 32\,GHz for both
sources, and at 15\,GHz for 3C\,223.1.

Analysis of the spectrum in different regions of the sources shows
remarkably little variation across the sources, and little evidence
for spectral curvature, even at the highest frequencies.  We fit a
theoretical spectrum to the data which corresponds to the largest
possible high frequency energy loss, and thereby obtain minimum
characteristic break frequencies.  Using these values and considering
the unavoidable Compton scattering off the microwave background, we
calculate firm upper limits on the particle ages of 34\,Myr and
17\,Myr.

From the analysis of the radio polarization properties, we conclude
that the sources do not reside in rich clusters. They probably reside
in similar environments to `classical' FRIIs of similar radio
power. The host galaxies are relatively undisturbed ellipticals.

From the details of the radio morphology we favour an explanation in
terms of a rapid realignment of the radio jet. Natarajan \& Pringle
(1998) have shown that the timescale for realignment of a viscous
accretion disc near a supermassive black hole could be sufficiently
short to explain these observations (less than a Myr for canonical AGN
parameters). The undisturbed properties of the host galaxies would
appear to rule out a merger with a large galaxy in the last $\sim
10^{8}$ year: considerably longer than the $\sim 10^7$yr since the jet
realignment.  We consider a binary black hole merger or aquisition of
a smaller galaxy as likely candidates for the cause of the change of
the jet axis.

There remain a number of unresolved questions, in particular:

(i) Do only sources in this small power range show `winged' structure?
All presently known examples are at similar radio power and
redshift. However this might be an observational selection effect, as
higher power/redshift sources might be expected to suffer greater
synchrotron and adiabatic losses and therefore more rapidly `lose'
their wings. In relation to this we wonder if the relatively flat
spectra of these two sources are integral or coincidental to their
morphology.

(ii) Why don't FRIs show evidence of such sudden realignments? This
cannot be attributed to a selection effect, but it may be that the
observed S- and Z-shaped structures occur instead when the jet axis
changes direction. If not, this must reflect either a difference in
their central regions, or a difference in propensity for the hosts to
capture, or contain, the required material with different angular
momentum.

(iii) Are the winged sources related to the double-double sources,
and, if so, how?  Is the cause of the realignment in the winged
sources related to the cause of the intermittent activity in
the double--doubles?

\subsection*{ACKNOWLEDGMENTS}
PAGS thanks Dr J.E Pringle for explaining to him the physical
processes leading to $\nu_2 >> \nu_1$.The National Radio Astronomy
Observatory is a facility of the National Science Foundation operated
under cooperative agreement by Associated Universities, Inc.  This
research was supported in part (JDT) by European Commission, TMR
Programme, Research Network Contract ERBFMRXCT96-0034 `CERES'

The remaining authors report with regret the death of their colleague 
Peter Scheuer on 21 Jan 2001. His deep insights into many 
astrophysical problems have been an inspiration to us all.

\bibliographystyle{mn}

\end{document}